\documentclass[pra,jmp,amsmath,amssymb,12pt,
reprint,showkeys,showpacs]{revtex4-1}
\usepackage{graphicx}
\usepackage{dcolumn}
\usepackage{bm}
\usepackage[mathlines]{lineno}
\usepackage{amsthm}
\usepackage[bookmarks]{hyperref}
\usepackage{pstricks}
\usepackage{dsfont}
\usepackage{mathrsfs}
\usepackage{breakurl}
\newtheorem{mydef}{Definition}[section]

\newtheorem{prop}{Proposition}[section]
\newtheorem{rem}{Remark}[section]

\begin{document}

\title[Some Remarks on  Quantum Brachistochrone]{Some Remarks on  Quantum Brachistochrone}

\author{F. Masillo}
\email{masillo@le.infn.it}
\affiliation{Dipartimento di Fisica,  Universit\`{a} del Salento}
\affiliation{INFN, Sezione di Lecce, I-73100 Lecce, Italy}
\date{\today}

\begin{abstract}
We study some aspects of the Quantum Brachistochrone Problem. Physical realizability of the faster pseudo Hermitian version of the problem is also discussed. This analysis, applied  to simple quantum gates,  supports an informational interpretation of the problem that is quasi Hermitian invariant.
\end{abstract}

\pacs{
03.65.-w,
03.65.Yz,
03.65.Ca,
03.67.-a,
03.67.Lx.
}
\keywords{}

\maketitle

\section*{Introduction}

Quantum computation is the theory studying the possibility to use quantum dynamical processes in solving computational problems \cite{NC00}. As in the standard computation case, one has  to quantify   the various costs (for example energy and time)  that the desired calculation  may require.
The search for optimal time evolutions, with limited resources, is a natural problem  that arises in such a context. The first part of this paper is just devoted to recall some results in literature.
In particular, in section \ref{sec.4.1} we give a precise mathematical formulation of this problem, the so called Quantum Brachistochrone,  and  expose  the solution  proposed in \cite{CHKO06}. In section  \ref{sec.4.2} we consider the same  problem in pseudo Hermitian quantum mechanics \cite{BBJM06} and the apparently paradoxical possibility of operating  a computational process in  an arbitrarily  small time and with limited energy costs \cite{M07}. Yet, section \ref{sec.4.3} clarifies the impossibility to produce a transition between Hermitian and pseudo Hermitian quantum mechanics \cite{M07d}.
Then, in the second part of the paper, we will try to overcome this no-go theorem  simulating a pseudo Hermitian dynamics by means of an Hermitian open dynamics (section   \ref{sec.4.4}). An alternative method, introduced in \cite{GS08} for PT symmetric Hamiltonians, is analyzed and generalized to pseudo Hermitian systems  in section \ref{sec.4.5}.

Carefully discussing both methods, however, we show clearly that a faster computation can only be obtained at the cost of expend more energy and/or of increase the uncertainty in the results of the computation.

Finally, in section \ref{sec.4.6}, some  further troublesome aspects of the Brachistochrone Problem, related to information theory, are examined, showing that nor a NOT gate nor a controlled NOT gate or a controlled $U$ gate can be realized in such way, without a relevant loss of efficiency.

\section{Quantum Brachistochrone}\label{sec.4.1}

The Quantum Brachistochrone Problem can be formulated as follows:

\emph{Given two quantum states $|\psi_i\rangle$ and $|\psi_f\rangle$, we want to find the (time-independent) Hamiltonian $H$ that performs the transformation
    \begin{equation}
    |\psi_i\rangle\rightarrow|\psi_f\rangle=e^{-\frac{i}{\hbar}H\tau}|\psi_i\rangle
    \end{equation}
in the minimal time $\tau$, for a fixed value of the  difference of eigenvalues of $H$.}

This problem was solved in \cite{CHKO06} where it was shown that, if we  put $\omega=|E_+- E_-|$, and
    \begin{equation}\label{s12}
    |\psi_i\rangle=\left(
    \begin{array}{c}
    1 \\
    0 \\
    \end{array}
    \right)\,\,\,\mbox{and}\,\,\,
    |\psi_f\rangle=\left(
    \begin{array}{c}
    a \\
    b \\
    \end{array}
    \right),\,\,\,(|a|^2+|b|^2=1),
    \end{equation}
the solution of the problem is given by the following Hermitian Hamiltonian:
    \begin{equation}\label{eq.hamilt}
    H=\left(
    \begin{array}{cc}
    s & \frac{\omega}{2}e^{-i\theta} \\
    \frac{\omega}{2}e^{i\theta} & s \\
    \end{array}
    \right),
    \end{equation}
where
\begin{equation}
    s=\frac{\omega \arg a}{2\arcsin|b|}
\end{equation}
and
\begin{equation}
    \theta=\arg (b)- \arg (a)-\frac{\pi}{2}.
\end{equation}
In particular if $a=|a|$ and $b=-i|b|$ we have that
    \begin{equation}\label{eq.hamilt2}
    H=\left(
    \begin{array}{cc}
    0 & \frac{\omega}{2} \\
    \frac{\omega}{2} & 0 \\
    \end{array}
    \right),
    \end{equation}
The minimal time to perform the required transformation is
    \begin{equation}\label{eq.tau}
    \tau=\frac{2\hbar}{\omega}\arccos|\langle\psi_i|\psi_f\rangle|=\frac{2\hbar}{\omega}\arccos|a|.
    \end{equation}
In what follows we will put $\hbar=1$.
Note that $\tau$ is always a strictly positive quantity (except the trivial case $|\psi_f\rangle=|\psi_i\rangle$), moreover it is inversely proportional to $\omega$. These  two observations make clear that to perform a faster transformation we need  to use as initial and final states two non orthogonal states or to increase $\omega$, i.e. we need more energy.
\newpage

\section{Ultrafast Pseudo-Hermitian Quantum Dynamics(?)}\label{sec.4.2}

We recall that \cite{M02c,M02b,M02a}
 \begin{mydef}
    A linear operator $H:{\mathscr H}\rightarrow{\mathscr H}$ is said pseudo Hermitian (or $\eta-$pseudo Hermitian) if and only if  an invertible linear hermitian operator $\eta$ exists such that
     \begin{equation}\label{eq.wpe}
        H^\dag=\eta H\eta^{-1}.
        \end{equation}
     Whenever $\eta>0$,  $H$ is said quasi Hermitian.
 \end{mydef}
 We recall further that the spectrum of a pseudo-Hermitian operator consists of complex conjugate pairs; in particular $H$ is  quasi Hermitian if and only if $H$ is diagonalizable with real spectrum \cite{SS03}. Note that,   for  a bidimensional pseudo Hermitian Hamiltonian $H$ if $(E_+-E_-)\in \mathds{R}\setminus\{0\}$ then $H$ is quasi Hermitian.

As it was shown in  \cite{BBJM06}, the  lower bound $\tau$ can be made arbitrarily low if, with the same eigenvalues limitations, we allow to use pseudo Hermitian Hamiltonians. This would  imply the realization of ultrafast computation with limited energy costs.

This  paradox can be explained using the isomorphisms existing between the different, but physically equivalent, quasi Hermitian descriptions of quantum mechanics \cite{M07b,M07}.

In particular, in the pseudo-Hermitian viewpoint the Hamiltonian operator is the key tool for the definition of a metric in the state space, hence, the lower bound for the time $\tau$ can be broken, for fixed states $|\psi_i\rangle$ and $|\psi_f\rangle$ because a metric  always  exists that reduces the distance between the given states. In other words, the use of a pseudo Hermitian Hamiltonian results in a reduction of the distance and consequently in a reduction of the time required for the transformation.

Let us consider for example, as distance measure the angle $A$ \cite{NC00}:
    \begin{mydef}
    The angle $A$ between two density matrices $\varrho$ and $\sigma$ is defined by:
        \begin{equation}\label{eq.angle}
        A(\varrho,\sigma)=\arccos  F(\varrho,\sigma)
        \end{equation}
    where $F$ is the fidelity:
        \begin{equation}
        F(\varrho,\sigma)=\mathrm{Tr} \sqrt{\varrho^\frac{1}{2}\sigma\varrho^\frac{1}{2}}.
        \end{equation}
    \end{mydef}
In particular the angle $A$ between the states in  (\ref{s12}) is
    \begin{equation}
    A=\arccos\mathrm{Tr}\sqrt{|\psi_i\rangle\langle\psi_i|\psi_f\rangle\langle\psi_f|
    \psi_i\rangle\langle\psi_i|}=\arccos|a|.
    \end{equation}
This last gives us the following
    \begin{equation}
    \tau=\frac{2}{\omega} A,
    \end{equation}
i.e., the velocity $v=\frac{\omega}{2}$ depends only on $\omega=|E_+-E_-|$.

In particular the angle between the states $|\psi_i\rangle$ and $|\psi_f\rangle$, with respect to the metric induced by a $\eta-$pseudo Hermitian Hamiltonian, is
    \begin{equation}
    A=\arccos\sqrt{\frac{\langle\psi_i|\eta|\psi_f\rangle\langle\psi_f|\eta|\psi_i\rangle}
    {\langle\psi_i|\eta|\psi_i\rangle\langle\psi_f|\eta|\psi_f\rangle}}.
    \end{equation}
By an appropriate choice of $\eta$ this distance can be made arbitrarily small, and in this way the time can be reduced to values near to zero. A possible choice is to put $\mathrm{det}\eta\rightarrow 0$, in fact when the determinant of a definite positive matrix $\eta$ goes to zero, the columns of $\eta$ become linearly dependent and so the initial and final states, with respect to the metric induced by $\eta$ become proportional, i.e. they coincide from a physical point of view.

This last observation means that if we could realize the passage from an Hermitian world to a pseudo Hermitian one we could reduce the transition time producing a ``wormhole effect'' in the state space.
It remains to explore if we can produce this ``wormhole effect'' in Hermitian quantum mechanics.

\section{Time dependent Pseudo-Hermitian Hamiltonian}\label{sec.4.3}

The problem posed at the end of the previous section  was first introduced in \cite{M07d},
    \footnote{This paper was the center of an interesting debate between Mostafazadeh and Znojil \cite{Z07a,M07a,Z07b,M07c}.}.
This problem can be synthetized in the following way.

Let us consider a system $S$. Let us suppose that the metric, that defines the geometry of the state space, is described by the time dependent operator $\eta(t)$. In particular
\begin{equation}
   \eta(t) =
   \left\{
     \begin{array}{cl}
       \mathbb{I} & \mbox{for}\,\,\, t\leq t_i, \\
       \eta'(t) &   \mbox{for}\,\,\,t_i<t< t_f , \\
       \tilde{\eta} & \mbox{for}\,\,\, t\geq t_f ,\\
     \end{array}
   \right.
\end{equation}
 We note that for $t\leq t_i$ the dynamics is Hermitian while for $t\geq t_f$ is $\tilde{\eta}-$pseudo Hermitian.

\emph{How can we describe the evolution of the system between   $t_i$ and $t_f$? }

We will call this  problem ``transition problem''.

The main assumption made in \cite{M07d} is that a Schr\"odinger equation still holds:
    \begin{equation}
    i \frac{\mathrm{d}}{\mathrm{d}t}U(t)=H(t)U(t).
    \end{equation}
To guarantee that $H(t)$ is an observable $\forall t$, it must  be  $\eta(t)$-pseudo Hermitian.
Applying these two assumptions we obtain that the evolution is ``unitary''(preserves the $\eta(t)-$norm) if and only if:
    \begin{equation}\label{}
    H^\dag(t)=\eta(t)H(t)\eta^{-1}(t)-i \eta(t)\frac{\mathrm{d}}{\mathrm{d}t}\eta^{-1}(t),
    \end{equation}
hence, $H(t)$ is $\eta(t)-$pseudo Hermitian if and only if
    \begin{equation}
    \eta(t)=\eta(0).
    \end{equation}
The previous equation implies that, in order to retain $H(t)$ interpretable as the energy operator during the evolution, we must give up the unitary of the process or $H(t)$ must be $\eta(t)$ pseudo Hermitian for a time independent $\eta$. This last possibility contradicts the assumption that $\eta(t_f)\neq\eta(t_i)$.

\section[PH Quantum Dynamics as Hermitian open Dynamics]{PH Quantum Dynamics as Hermitian open Dynamics}\label{sec.4.4}

In order to overcome the negative answer to the transition problem,
in this section we will try to simulate a pseudo-Hermitian dynamics as an Hermitian open quantum dynamics.
This point of view will induce the paradoxically consequence that to obtain a faster evolution, under the same  eigenvalues limitations we have to use a dissipative process.
This apparent contradiction shows that the eigenvalues of a pseudo Hermitian Hamiltonian cannot always be interpreted as  energy values. We will clarify our viewpoint at the end of the section.

Let us consider a bidimensional Hilbert space ${\mathscr H}$.
We recall that a generic $\eta$-quasi Hermitian Hamiltonian $H:{\mathscr H}\rightarrow{\mathscr H}$ can always be written in the form:
    \begin{equation}
    H=\eta^{-\frac{1}{2}}h\eta^{\frac{1}{2}},
    \end{equation}
for some Hermitian operator $h:{\mathscr H}\rightarrow{\mathscr H}$.

Let us consider  an orthonormal basis such that $\eta$ is in diagonal form, and
without loss of generality we can suppose that
    \begin{equation}
    \eta=\left(
    \begin{array}{cc}
    1 & 0 \\
    0 & \lambda^2 \\
    \end{array}
    \right),
    \end{equation}
(in fact if $H$ is $\eta$-quasi Hermitian then it is $\alpha\eta$-quasi Hermitian for all $\alpha \in\mathds{R}$).
Moreover we can suppose that
    \begin{equation}\label{eq.4.17}
    h=\left(
    \begin{array}{cc}
    a & d \\
    d & c \\
    \end{array}
    \right),
    \end{equation}
with $a,c,d\in \mathds{R}$ (if $d=|d|e^{i\theta}$, we can operate a change of basis by the unitary matrix $\mathrm{diag}(1,e^{-i\theta})$). So we can consider only quasi Hermitian Hamiltonian of the form:
    \begin{equation}\label{eq.Ham}
    H=\left(
    \begin{array}{cc}
    a & \lambda d \\
    \lambda^{-1}d & c \\
    \end{array}
    \right),
    \end{equation}
with $\lambda>0$.

Let us now consider the standard decomposition of a matrix in Hermitian and anti Hermitian part:
    \begin{equation}
    H=H_1+iH_2,
    \end{equation}
where $H_1,H_2$ are Hermitian matrix. We have:
    \begin{equation}
    H_1=\left(
    \begin{array}{cc}
    a & \frac{\lambda +\lambda^{-1}}{2}d \\
    \frac{\lambda +\lambda^{-1}}{2}d & c \\
    \end{array}
    \right)
    \end{equation}
and
    \begin{equation}\label{tleq}
    H_2=\left(
    \begin{array}{cc}
    0 & \frac{\lambda -\lambda^{-1}}{2i}d \\
    \frac{-\lambda +\lambda^{-1}}{2i}d & 0 \\
    \end{array}
    \right)
    \end{equation}

  Note that  $H_2$ is a definite matrix, the null matrix,  if and only if $H$ is Hermitian, i.e. if $d=0$ or $\lambda=1$.

Moreover    $H_2$ is a traceless matrix, so the eigenvalues of $H_2$ $\mu_1$ and $\mu_2(\leq\mu_1)$ satisfy the following relation:
        \begin{equation}
        \mu_1=-\mu_2=\frac{|\lambda -\lambda^{-1}|}{2}|d|.
        \end{equation}

Let us now consider the semigroup generated by $H$, where $H$ is a quasi Hermitian operator.
We have  the following equation:
    \begin{equation}
    \varrho(t)=e^{-iHt}\varrho e^{iH^\dag t}.
    \end{equation}
Differentiating both sides of the previous equation we obtain a Lindblad-Kossakowsky  type equation:
    \begin{equation}\label{eq.LKT}
    \frac{\mathrm{d}}{\mathrm{d}t}\varrho=-i[H_1,\varrho]+\{H_2,\varrho\}.
    \end{equation}
    Note that if $H_2=0$ the previous equation reduce to the Liouville-von Neumann equation, so it is natural to call $-i[H_1,\cdot]$ hamiltonian term.

It is simple to show that:
    \begin{equation}
    \frac{\mathrm{d}}{\mathrm{d}t}\mathrm{Tr}(\varrho)
    =\mathrm{Tr}\{H_2,\varrho\}=2\mathrm{Tr}(H_2\varrho).
    \end{equation}
As simple consequence we obtain that:
    \begin{prop}
    The semigroup generated by a quasi Hermitian Hamiltonian  is not trace preserving. In particular it is not strictly trace decreasing or strictly trace increasing.
    \end{prop}

Then a non  trivial problem arises. In fact if we accept the probabilistic interpretation of the density matrix we cannot give a well defined interpretation of probabilities greater than one.

A simple way to overcome these difficulties is to normalize by hand the resulting density operator. This solution appears problematic because it introduces a non linear operation that cannot be extended to arbitrary density operator.
In fact the greater obstacle to non linear evolution in quantum theory is that it produces different evolutions for different decompositions of the same density operator (this last problem is not so problematic if we consider only pure states, because they admit a unique decomposition).

Another way to solve this problem is to introduce a global  dissipative term in the Hamiltonian:
    \begin{equation}
    H'=H-iD,
    \end{equation}
with $D=\mu_1 \mathbb{I}$, where $\mathbb{I}:{\mathscr H}\rightarrow{\mathscr H}$  is the identity operator on ${\mathscr H}$, and $\mu_1 $ is the maximal eigenvalue of $H_2$.
In this way the density matrices obtained using  $H'$ as generator of the semigroup:
    \begin{enumerate}
        \item are proportional to the original ones, i.e.
            \begin{equation}
            e^{-iHt}\varrho e^{+iHt}=ke^{-iH't}\varrho e^{+iH't},
            \end{equation}
    with $k=e^{-2\mu_1t}$;
        \item $\mathrm{Tr}(\varrho(t))\leq1$ for all $t\geq0$.
    \end{enumerate}
Note that $D=\mu_1 \mathbb{I}$ is the minimal dissipative term that satisfies the previous conditions.

In particular because the dissipative term $D$ is a multiple of the identity $\mathbb{I}$ we can separate the process in a deterministic evolution governed by the $\eta-$quasi Hermitian Hamiltonian $H$ and a purely dissipative process caused by $D$.
The minimal time $\tau$ to operate the transformation
    \begin{equation}\label{eq.tr}
    |\psi_i\rangle \rightarrow |\psi_f\rangle,
    \end{equation}
can be obtained by the analysis of the equation:
    \begin{equation}\label{eq.tr}
    |\psi_i\rangle \rightarrow |\psi_f\rangle=e^{-iHt}|\psi_i\rangle,
    \end{equation}
where $H=\eta^{-\frac{1}{2}}h\eta^{\frac{1}{2}}$.
This transformation can be written in the form:
    \begin{equation}\label{eq.HEE}
    |\psi_i'\rangle\rightarrow |\psi_f'\rangle=e^{-iht}|\psi_i'\rangle,
    \end{equation}
where
\begin{equation}
|\psi_i'\rangle=\frac{\eta^{\frac{1}{2}}|\psi_i\rangle}{\sqrt{\langle\psi_i\eta|\psi_i\rangle}} \,\,\,\mbox{and}\,\,\,
|\psi_f'\rangle=\frac{\eta^{\frac{1}{2}}|\psi_f\rangle}{\sqrt{\langle\psi_f\eta|\psi_f\rangle}}.
\end{equation}
Note that the transformation (\ref{eq.HEE}) is a unitary one generated by the Hermitian operator $h$, so to obtain the minimal time to perform the transformation (\ref{eq.HEE}) we can use the results obtained in \cite{BBJM06}:
the minimal time $\tau$ to perform the transformation
    \begin{equation}
    |\psi_i'\rangle=\left(
    \begin{array}{c}
    1 \\
    0 \\
    \end{array}
    \right)\rightarrow
    |\psi_f'\rangle=\left(
    \begin{array}{c}
    a \\
    b \\
    \end{array}
    \right),
    \end{equation}
with $h$ Hermitian operator with fixed $\omega$, is
    \begin{equation}\label{ea.MTE}
    \tau=\frac{2}{\omega}\arccos|a|.
    \end{equation}

To quantify the effect of the dissipative term, we operate a  basis transformation, so we can consider
\begin{equation}
\eta^\frac{1}{2}=\left(
\begin{array}{cc}
1 & g \\
g^* & f \\
\end{array}
\right),
\end{equation}
where $f\in\mathbb{R}$ and $f-gg^*>0$ (we recall that $(\eta^\frac{1}{2})_{11}$ can be always posed equal to $1$ multiplying  $\eta^\frac{1}{2}$ by an irrelevant factor $\alpha\in \mathds{R}$).

Then
    \begin{equation}
    \eta^{-\frac{1}{2}}=\frac{1}{f-gg^*}\left(
    \begin{array}{cc}
    f & -g \\
    -g^* & 1 \\
    \end{array}
    \right),
    \end{equation}
and
    \begin{equation}
    \eta=\left(
    \begin{array}{cc}
    1+gg^* & g(1+f) \\
    g^*(1+f) & f^2-gg^* \\
    \end{array}
    \right).
    \end{equation}
As particular case, let us consider two orthogonal states
    \begin{equation}
    |\psi_i\rangle=\left(
    \begin{array}{c}
    1 \\
    0 \\
    \end{array}
    \right)\,\,\,\mbox{and}\,\,\,
    |\psi_f\rangle=\left(
    \begin{array}{c}
    0 \\
    1 \\
    \end{array}
    \right).
    \end{equation}
Recalling equation (\ref{eq.HEE}), we need to find
    \begin{equation}
    |\psi_i'\rangle=\frac{\eta^{\frac{1}{2}}|\psi_i\rangle}{\sqrt{\langle\psi_i\eta|\psi_i\rangle}} \,\,\,\mbox{and}\,\,\,
    |\psi_f'\rangle=\frac{\eta^{\frac{1}{2}}|\psi_f\rangle}{\sqrt{\langle\psi_f\eta|\psi_f\rangle}}.
    \end{equation}
So we have:
    \begin{equation}
    |\psi_i'\rangle=
    \frac{1}{\sqrt{1+gg^*}}\left(
    \begin{array}{c}
    1 \\
    g^* \\
    \end{array}
    \right)
    \,\,\,\mbox{and}\,\,\,
    |\psi_f'\rangle=
    \frac{1}{\sqrt{f^2+gg^*}}\left(
    \begin{array}{c}
    g \\
    f \\
    \end{array}
    \right)
    \end{equation}
In particular we have that
\begin{equation}\label{eq.4.39}
|a'|=\sqrt{|\langle\psi_i'|\psi_f'\rangle|^2}=\sqrt{1-\frac{(f-gg^*)^2}{(1+e e^*) (f^2+gg^*)}}.
\end{equation}

Operating a basis transformation, represented by the unitary matrices $U^\dag$, where
\begin{equation}
    U=\frac{1}{\sqrt{1+gg^*}}\left(
    \begin{array}{cc}
    1 & g \\
    g^* & -1 \\
    \end{array}
    \right)\left(
         \begin{array}{cc}
            e^{i\gamma_1} & 0 \\
           0 & e^{i\gamma_2} \\
         \end{array}
       \right)
    \end{equation}
we can obtain, by an appropriate choice of $\gamma_1$ and $\gamma_2\in\mathds{R}$:
\begin{equation}
   |\psi_i'\rangle=\left(
                     \begin{array}{c}
                       1 \\
                       0 \\
                     \end{array}
                   \right)
                   \,\,\,\mbox{and}\,\,\,
   |\psi_f'\rangle=\left(
                     \begin{array}{c}
                       a' \\
                       ib' \\
                     \end{array}
                   \right)
\end{equation}
with $a',b'\in\mathds{R}^+$.

\begin{figure}
  \includegraphics[width=6cm]{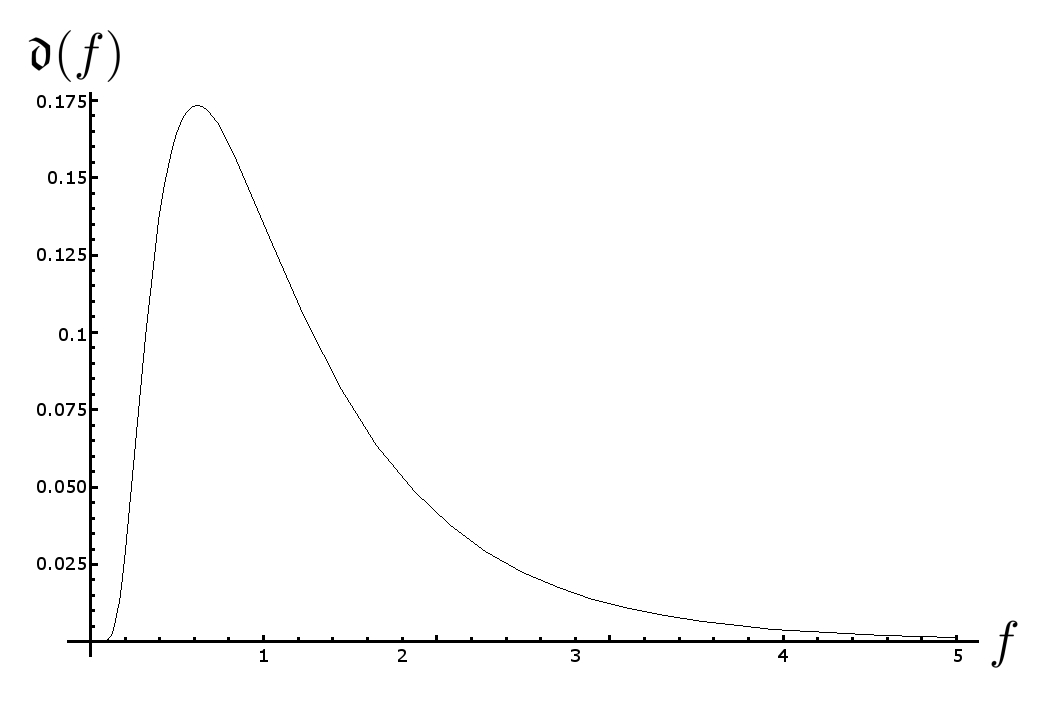}\\
  \caption{The dissipative factor $\mathfrak{d}(f)$}\label{fig.1.1}
\end{figure}

The right $\eta-$pseudo Hermitian Hamiltonian   is then given, in the original basis, by
    \begin{equation}
    h=U h'U^\dag,
    \end{equation}
where, using equation (\ref{eq.hamilt2}),
 \begin{equation}
    h'=\left(
    \begin{array}{cc}
    0 & \frac{\omega}{2} \\
    \frac{\omega}{2} & 0 \\
    \end{array}
    \right).
    \end{equation}
 The original Hamiltonian $H$ has then the form:
    \begin{align}
    H=&\eta^{-\frac{1}{2}}Uh'U^\dag\eta^{\frac{1}{2}}=\nonumber\\
    =&\frac{\omega}{2}\left(
         \begin{array}{cc}
            e^{-i\gamma_1} & 0 \\
           0 & e^{-i\gamma_2} \\
         \end{array}
       \right)\nonumber\\&\left(
    \begin{array}{cc}
    \frac{e (1+f)}{f-g'g'^*} & 1+\frac{g^2 (1+f)}{f-g'g'^*}+\frac{(-1+g^2) (1+f)}{1+g'g'^*}\\
    1+\frac{1+f}{-f+g'g'^*} & \frac{e (1+f)}{-f+g'g'^*} \\
    \end{array}
    \right)\nonumber\\
    &\left(
         \begin{array}{cc}
            e^{i\gamma_1} & 0 \\
           0 & e^{i\gamma_2} \\
         \end{array}
       \right)
    \end{align}
where $g'=ge^{i(\gamma_2-\gamma_1)}$.
Then, let us consider $H_1=\frac{H+H^\dag}{2}$ and $H_2=\frac{H-H^\dag}{2i}$.
Note that $H_1$ is a traceless matrix \footnote{$H$ is a traceless matrix, $H_2$ is a traceless matrix (see equation (\ref{tleq})) and so we have the thesis.}
and so has two symmetric eigenvalues $E_+=E$ and $E_-=-E$.
In particular we obtain that in the limit $\tau\rightarrow 0$, i.e. $f\rightarrow gg^*$, we have that
\begin{equation}\label{equ.1}
(2E)^2=(E_+-E_-)^2=(\mathrm{Tr} H_1)^2-4\mathrm{Det} H_1\rightarrow \infty.
\end{equation}
Moreover the analysis of the dissipative factor $\mathfrak{d}(g,f)$  gives, in the limit $gg^*\rightarrow f $, the following  results:
\begin{align}
\mathfrak{d}(f)=\lim_{gg^*\rightarrow f}\mathfrak{d}(g,f)=\lim_{gg^*\rightarrow f}\frac{\langle\psi_f|\psi_f\rangle}{\langle\psi_i|\psi_i\rangle}=&\\
=\lim_{gg^*\rightarrow f}\frac{\langle\psi_i|e^{-i(H-H^\dag)\tau}\psi_i\rangle}{\langle\psi_i|\psi_i\rangle}
=&\frac{1}{f}e^{-(\frac{1}{f}+f)}.
\end{align}
A graphical representation of the dissipative factor $\mathfrak{d}$ vs. $f$ is given in figure \ref{fig.1.1}.

We can observe that the dynamics is strongly dissipative (the probability of revelation is reduced to values less than $20\%$).
Equation (\ref{equ.1})   justifies a posteriori the usual identification of $H_1$ with the Hamiltonian of the system and its eigenvalues can be considered as energy values.
In fact if we want realize the transformation generated by the Hamiltonian $H$ as an open dynamics we implicitly assume that an auxiliary system $A$ exists  such that $S+A$ is a closed systems governed by the Hermitian Hamiltonian $H^{S+A}$. Now, in general, an open dynamics is the reduced dynamics of a broader open system so the time to operate the transformation (\ref{eq.tr}), performed by the dynamical semigroup generated by $H$ is the same as the time to perform the closed evolution
\begin{equation}\label{eq.tr2}
   |\psi^{(S)}_I\rangle\otimes |\phi^{(A)}_I\rangle \rightarrow |\psi^{(S)}_F\rangle\otimes |\phi^{(A)}_F\rangle,
\end{equation}
by a new Hermitian operator $H^{S+A}$.
In particular under the energy constraints discussed, this time can not be made arbitrarily small because  the distance between the initial and final states in the compound system $S+A$ is always greater than the distance between  the reduced initial and final states (the distance  cannot increase by  operating a partial trace operation \cite{NC00}).

To conclude, in this section we shown that the pseudo Hermitian evolution dynamics can be obtained by means of an open dynamics in the Hermitian world, but such process is strongly dissipative, so that all this  has an energy cost.
The same argument holds if we use instead of a pseudo Hermitian Hamiltonian $H$ a more general non Hermitian Hamiltonian subjected to the only restriction that the difference of the eigenvalues is real and fixed (see \cite{AF08},\footnote{In this paper we show that the faster then Hermitian evolution of non Hermitian dynamics is obtained increasing the energy cost, while in \cite{AF08} this aspect was neglected.}).

\section[PH Quantum Dynamics as Hermitian open Dynamics II]{PH Quantum Dynamics as Hermitian\\ open Dynamics II}\label{sec.4.5}

In this section we analyze a different method presented in \cite{GS08}, and here  generalized to quasi Hermitian Hamiltonians $H$, in order to simulate a pseudo-Hermitian evolution by means of a Hermitian open dynamics.
As we said above, any $\eta-$quasi Hermitian Hamiltonian $H:{\mathscr H}^2\rightarrow{\mathscr H}^2$ can be written as
    \begin{equation}
    H=\eta^{-\frac{1}{2}}h\eta^{\frac{1}{2}},
    \end{equation}
where $h=h^\dag$. Without loss of generality, we can put  $\mathrm{Tr}h=0$ (this last condition relies on the fact that the only important quantity is the difference of the eigenvalues $\omega$).
The aim of reference \cite{GS08}  was to provide a way to obtain the pseudo Hermitian evolution in ${\mathscr H}^2$  starting from the Hermitian dynamics of a larger system ${\mathscr H}^2\oplus\tilde{{\mathscr H}}^2$,  i.e.  realizing
the following transformation:
    \begin{equation}\label{eq.2.49}
    \bm\phi_i\rightarrow  \bm{\phi}(t)=\mathbf{U}(t)\bm{\phi}_i,
    \end{equation}
where $\mathbf{U}(t)$ is a unitary transformation (generated by a Hermitian Hamiltonian $\mathbf{H}$), and
    \begin{equation}
    \bm{\phi}(t)=
        \left(
        \begin{array}{c}
        \psi(t) \\
        \chi(t) \\
        \end{array}
        \right),
    \end{equation}
with the requirement that
    \begin{equation}\label{equ222}
    \psi(t)=e^{-iH}\psi_i,
    \end{equation}
for all $t\geq0$. In other words, the state vector $|\psi\rangle$ represents only partially the state of the system. The unobserved part, represented by $\chi$ lies in the space $\tilde{{\mathscr H}}^2$.  We will call $\tilde{{\mathscr H}}^2$ unobserved space and its elements unobserved states.

To realize equation (\ref{eq.2.49}), let us consider the eigenvectors of $H$ and $H^\dag$; in particular, if we denote with $|e_+\rangle$ and $|e_-\rangle$ the (normalized) eigenvector of $h$ relative to the eigenvalues $\frac{\omega}{2}$ and $-\frac{\omega}{2}$ respectively, we have immediately
    \begin{equation}
     H\eta^{-\frac{1}{2}}|e_\pm\rangle=\pm \frac{\omega}{2}\eta^{-\frac{1}{2}}|e_\pm\rangle
    \end{equation}
and
    \begin{equation}
    H^\dag\eta^{\frac{1}{2}}|e_\pm\rangle=\pm \frac{\omega}{2}\eta^{\frac{1}{2}}|e_\pm\rangle
    \end{equation}
The non orthonormal set $$\{\eta^{-\frac{1}{2}}|e_+\rangle,\eta^{-\frac{1}{2}}|e_-\rangle,\eta^{\frac{1}{2}}|e_+\rangle,\eta^{\frac{1}{2}}|e_-\rangle\}$$
can be transformed into an orthonormal set if we dilate the Hilbert space ${\mathscr H}^2$ into ${\mathscr H}^2\oplus\tilde{{\mathscr H}}^2$ by the so called Naimark dilation.
\begin{rem}
In the subsequent discussion we use as bases for the matrix representation in ${\mathscr H}^2$ the orthonormal set $\{|e_+\rangle, |e_-\rangle\}$. In particular the column of $\eta^{-\frac{1}{2}}(\eta^{\frac{1}{2}})$ are the eigenvectors of $H $ ($H^\dag$ respectively) and then we have:
\begin{equation}\label{eq.4.52}
H\eta^{-\frac{1}{2}}=\eta^{-\frac{1}{2}}\tilde{E}
\end{equation}
and
\begin{equation}\label{eq.4.53}
H^\dag\eta^{\frac{1}{2}}=\eta^{\frac{1}{2}}\tilde{E},
\end{equation}
where $\tilde{E}=\mathrm{diag}(\frac{\omega}{2},-\frac{\omega}{2})$.
\end{rem}
Denoting by ${\bm V}=[|v_1\rangle,|v_2\rangle,|v_3\rangle,|v_4\rangle]$ the matrix of the extended vectors we have
    \begin{equation}
    {\bm V}=f
    \left(
    \begin{array}{cc}
    \eta^{-\frac{1}{2}} & \eta^{\frac{1}{2}} \\
    X & Y \\
    \end{array}
    \right),
    \end{equation}
where $f$ is an normalization factor and $X,Y$ are $2\times 2$ matrices.
Imposing the orthonormalization of the column of ${\bm V}$ it is simple to show that $\mathrm{det}\eta=1$ (this condition can always be supposed following similar arguments following equation (\ref{eq.4.17}))
$X=\eta^{\frac{1}{2}}$ and $Y=-\eta^{-\frac{1}{2}}$, $f=\frac{1}{\sqrt{\mathrm{Tr}(\eta)}}$.
To obtain the matrix $\bm H$ let us impose that it is possible to recover  the original model when we restrict to the first two row of $\bm V$ in the new model. In particular from equation (\ref{eq.4.52}), (\ref{eq.4.53}) and an ansatz
$f [H\eta^{-\frac{1}{2}},H^\dag\eta^{\frac{1}{2}}]=f [\eta^{-\frac{1}{2}},\eta^{\frac{1}{2}}]\bm E$ where
$\bm E=(\frac{\omega}{2},-\frac{\omega}{2},\frac{\omega}{2},-\frac{\omega}{2})$, it is possible to show that
    \begin{equation}
    \mathbf{H}=f^2\left(
    \begin{array}{cc}
    H\eta^{-1}+\eta H & H-H^\dag \\
    H^\dag-H & H\eta^{-1}+\eta H \\
    \end{array}
    \right),
    \end{equation}
where $f=\sqrt{\frac{\cos \alpha}{2}}$.
In order to reproduce equation (\ref{equ222})the initial state must be chosen as
\begin{equation}
    \bm{\phi}_i=
        \left(
        \begin{array}{c}
        \psi_i \\
        \chi_i \\
        \end{array}
        \right)
        =
        \left(
        \begin{array}{c}
        \psi_i \\
        \eta  \psi_i \\
        \end{array}
        \right).
    \end{equation}
In particular for this initial choice the time evolution of the system is described by
\begin{equation}
    \bm{\phi}(t)={\bm U }(t)\bm{\phi}_i=
    \left(
      \begin{array}{cc}
        U(t) & 0 \\
        0 & \eta U(t)\eta^{-1} \\
      \end{array}
    \right)
        \left(
        \begin{array}{c}
        \psi_i \\
        \chi_i \\
        \end{array}
        \right).
   \end{equation}
The subsequent discussion in \cite{GS08} makes evident that the procedure used by the authors is a method to reduce  the distance between the starting  and final states, in particular in the limit  $\tau=0$ the two states appear indistinguishable.

In fact by the results contained in the previous section,  $\tau$ vanishes when the columns of $\eta$ becomes linearly independent  (see equation (\ref{eq.4.39})).
In particular this means that to preserve  condition $\mathrm{det}\eta =1$ we have to substitute   $\eta$ with $\eta'=\frac{1}{{\mathrm{det}\eta}}\eta$; then,
$\mathrm{det}\eta\rightarrow0$ implies $\frac{\langle\psi_i|\psi_i\rangle}{\langle\chi_i|\chi_i\rangle}\rightarrow0$, so that the dominant term in $\bm{\phi}_i$ is the hidden state $|\chi_i\rangle$. Analogous conclusion holds for $\bm{\phi}_f$. Moreover, in this limit, $|\chi_i\rangle\approx|\chi_f\rangle$ and so the two states becomes indistinguishable.

We note that this last remark agrees with our previous discussions at the end of section \ref{sec.4.1}.

In conclusion, the  methods proposed in this section and in section \ref{sec.4.4} share the nice feature to solve the transition problem between Hermitian and pseudo Hermitian evolution presented in section \ref{sec.4.3} (in fact the evolution is governed by the Hermitian Hamiltonian $\bm H$). Yet, as it is evident, the advantage in time produced by such methods is deleted by the introduction of a dissipative effect or by a reduction in the probability of revelation of the final state.

\section{Some remarks on information theory}\label{sec.4.6}

As already said in the introduction, in \cite{BBJM06}, the authors suggest that pseudo Hermitian Hamiltonians can be used to realize ultrafast computation  with limited energy cost (in particular as a way of carrying out a NOT gate, i.e., a transformation between orthogonal states).

According our previous analysis this ultrafast  computation can be realized only in two ways:
    \begin{itemize}
    \item increasing the difference of the energy values (see section \ref{sec.4.4});
    \item reducing the distance between the initial and final states (see section \ref{sec.4.5}).
    \end{itemize}

This last possibility can be called ``Computation over a non orthogonal computational basis''.

In this section we will point out some troublesome implications of such computation in information theory.

\subsection{Computation not interpretable}

In this subsection we analyze some kinematical aspects on the use of non orthogonal computational bases.
Let us consider the bidimensional case, and choose as computational basis $\mathcal{B}$ the states
    \begin{equation}
    |\psi_0\rangle=
    \left(
    \begin{array}{c}
    1 \\
    0 \\
    \end{array}
    \right)
    \,\,\,\mbox{and}\,\,\,
    |\psi_0\rangle=\left(
    \begin{array}{c}
    a \\
    b \\
    \end{array}
    \right),
    \end{equation}
with $|a|^2+|b|^2=1$.
A possible Positive Operator Valued Measure \cite{NC00} that can be used to establish the results of the computation in this basis is composed by the operator elements:
    \begin{equation}
    E_0=\left(\begin{array}{c}
    b^* \\
    -a^* \\
    \end{array}\right)\left(
    \begin{array}{cc}
    b & -a \\
    \end{array}
    \right)=\left(
    \begin{array}{cc}
    |b|^2 & -ab^* \\
    -a^*b & |a|^2\\
    \end{array}
    \right),
    \end{equation}
    \begin{equation}
    E_1=
    \left(\begin{array}{c}
    0 \\
    1 \\
    \end{array}\right)
    \left(
    \begin{array}{cc}
    0 & 1 \\
    \end{array}
    \right)=\left(
    \begin{array}{cc}
    0 & 0 \\
    0 & 1\\
    \end{array}
    \right),
    \end{equation}
and
    \begin{equation}
    E_2=\mathbb{I}-\frac{1}{1+|a|}(E_0+E_1),
    \end{equation}
associated respectively to the results $0,1,2$: if the result of the measure is $0 (1)$, the final states can only be
$|\psi_0\rangle$ (respectively $|\psi_1\rangle$), while if we obtain as result the value $2$ we cannot discriminate between the two possible final states.
In particular the probability  $\mathcal{P}_{|\psi_i\rangle}(2)$ to obtain as result $2$ when the input state is  $|\psi_i\rangle$ is:
    \begin{equation}\label{eq.4.60}
    \mathcal{P}_{|\psi_1\rangle}(2)=\mathcal{P}_{|\psi_2\rangle}(2)=|a|.
    \end{equation}
This last represents the probability that the result of the  computation cannot be interpreted.
Comparing equation (\ref{eq.4.60}) with the time $\tau$ in equation (\ref{ea.MTE}):
    \begin{equation}
    \tau=\frac{2}{\omega}\arccos|a|,
    \end{equation}
we see that if we decrease the time $\tau$ needed to operate the transformation $|\psi_0\rangle\rightarrow|\psi_1\rangle$, we increase the probability to obtain a non interpretable  result.

\subsection{Quantum NOT-Gate}

As it was said above, some authors suggest to apply the Brachistochrone  Problem  in the construction of quantum NOT-gates. To increase the speed of the computation we agree to use a non orthogonal computational basis:
\begin{equation}
  |\psi_0\rangle, |\psi_1\rangle,\,\,\, \langle\psi_0|\psi_1\rangle\neq0.
\end{equation}
Without loss of generality, we can put
\begin{equation}\label{eq.st.}
    |\psi_0\rangle=
    \left(
    \begin{array}{c}
    1 \\
    0 \\
    \end{array}
    \right)
    \,\,\,\mbox{and}\,\,\,
    |\psi_0\rangle=\left(
    \begin{array}{c}
    \cos \frac{\theta}{2} \\
    -i\sin\frac{\theta}{2} \\
    \end{array}
    \right)
    \end{equation}
So the state
$|\psi_1\rangle$ is identified in the Bloch sphere by the angles $\theta$ and $\phi=\frac{\pi}{2}$.
Let us suppose that the transformation is generated by an Hermitian Hamiltonian  $H$.
If we require that, after a time $\tau$ the transformation $e^{-iH\tau}$ produces the desired transformations:
\begin{equation}\label{tr.1}
  |\psi_0\rangle\rightarrow |\psi_1\rangle,
\end{equation}
and
\begin{equation}\label{tr.2}
  |\psi_1\rangle\rightarrow |\psi_0\rangle,
\end{equation}
we obtain that the faster hamiltonian $H$ that realize the transformation (\ref{tr.1}) doesn't realize (\ref{tr.2}), unless  $\langle\psi_0|\psi_1\rangle=0$.
In fact  it is simple to show that for the states (\ref{eq.st.}), we must have:
\begin{equation}\label{tr5}
    e^{-iH\tau}=\left(
                                \begin{array}{cc}
                                  \cos\frac{\theta}{2} & -i \sin\frac{\theta}{2} \\
                                   -i \sin\frac{\theta}{2} &  \cos\frac{\theta}{2} \\
                                \end{array}
                              \right)
\end{equation}
and then
\begin{align}\label{tr.3}
    e^{-iH\tau}|\psi_1\rangle=\left(
                                \begin{array}{cc}
                                  \cos\frac{\theta}{2} & -i \sin\frac{\theta}{2} \\
                                   -i \sin\frac{\theta}{2} &  \cos\frac{\theta}{2} \\
                                \end{array}
                              \right)\left(
    \begin{array}{c}
    \cos \frac{\theta}{2} \\
    -i\sin\frac{\theta}{2} \\
    \end{array}
    \right)&\nonumber
    \\=\left(
    \begin{array}{c}
    \cos \theta \\
    -i\sin\theta \\
    \end{array}
    \right)\neq|\psi_0\rangle&
    \end{align}

More generally, this dynamical problem, connected to the use of non orthogonal computational bases, puts a natural question:

\emph{ Is there an Hermitian Hamiltonian $H$ that realizes the $NOT$ gate in this non orthogonal basis in a minimal time $\tau$?}

Using the Bloch sphere representation,  \cite{GR08}, it is possible to show that
an arbitrary unitary evolution corresponds to a rotation in the Bloch sphere. Moreover, if we suppose that some unitary transformation $U(\tau)$  realizes the transformations (\ref{tr.1}) and (\ref{tr.2}) in the  time $\tau$, then the unitary  transformation $U(2\tau)$ is the identity transformation and we can conclude that
\begin{equation}
   \frac{ \omega}{2}2\tau=\pi.
\end{equation}
Comparing this last with the time needed by the faster Hamiltonian that produces the transformation $|0\rangle\rightarrow|1\rangle$ (we need only to put $\theta=\pi$) we obtain
that the use of a nonorthogonal base doesn't produce any advantages.

\begin{figure}
  \includegraphics[width=5cm]{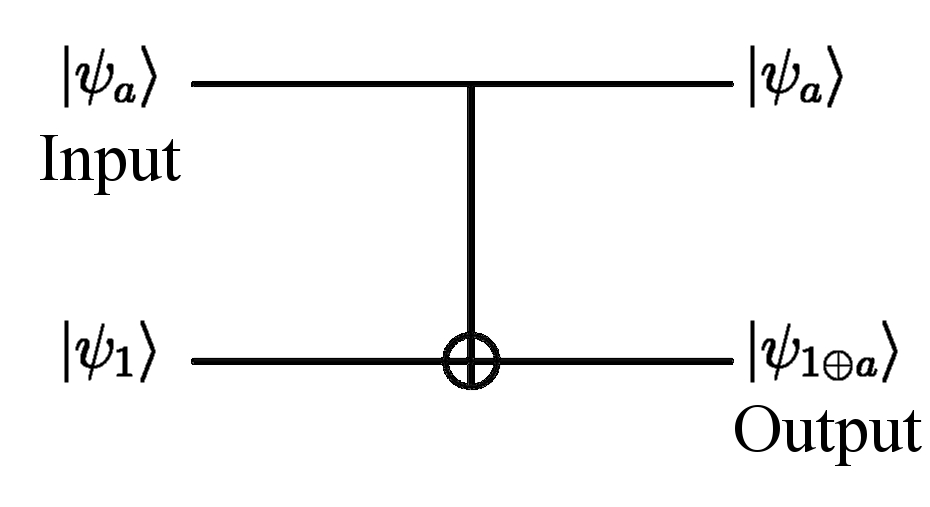}\\
  \caption{A Control Not type gate for non orthogonal basis. ($a=0,1$; $1\oplus a$ denotes a sum \texttt{mod} $2$).}\label{Untitled}
\end{figure}

One can try to overcome this last problem using a controlled NOT gate (see figure \ref{Untitled}).
 As it is evident the action of this gate is to perform the transformation $|\psi_1\rangle\rightarrow |\psi_0\rangle$ in the time $\tau$, while $ |\psi_0\rangle\rightarrow |\psi_1\rangle$ is automatically performed (in a time $\tau=0$).
As it is simple to note this last corresponds to the total transformations:
\begin{equation}\label{eq.cnot}
   \begin{array}{c}
    |\psi_1\rangle|\psi_1\rangle\rightarrow|\psi_1\rangle|\psi_0\rangle \\
     |\psi_0\rangle|\psi_1\rangle\rightarrow|\psi_0\rangle|\psi_1\rangle
   \end{array}
\end{equation}
Such transformations, however, are not realizable by a unitary transformation. Indeed  the validity of (\ref{eq.cnot}) appears strictly connected to the no cloning theorem \cite{NC00}, i.e. the impossibility to copy non orthogonal states using completely positive maps.
In fact by an appropriate  unitary transformation acting only on the first (or second) subsystem it would be  possible to realize the following transformations:
\begin{equation}\label{eq.cnot2}
   \begin{array}{c}
    |\psi_1\rangle|\psi_1\rangle\rightarrow|\psi_1\rangle|\psi_1\rangle, \\
     |\psi_0\rangle|\psi_1\rangle\rightarrow|\psi_0\rangle|\psi_0\rangle,
   \end{array}
\end{equation}
that is to copy the first qubit. Hence, also the possibility of realizing a controlled NOT gate is excluded. Finally we can try to use a standard control$-U$  operation of the form
\begin{equation}
  V=|e_1\rangle\langle e_1|\otimes U+ |e_0\rangle\langle e_0|\otimes \mathds{I},
\end{equation}
with $\langle e_0|e_1\rangle=0$ and $U=e^{-i\tau H}$ with $H$ given in  equation (\ref{eq.hamilt}). It is simple to prove that the resulting quantum operation
 is
 \begin{equation}
\varepsilon[\varrho_{\mathrm{Input}}]=\mathrm{Tr}_{\mathrm{Input}}(V\varrho_{\mathrm{Input}}\otimes|e_1\rangle\langle e_1|V^\dag).
 \end{equation}
In particular we obtain $$\varepsilon[|\psi_1\rangle\langle\psi_1|]=p|\psi_0\rangle\langle\psi_0|+(1-p)|\psi_1\rangle\langle\psi_1|,$$
and
$$\varepsilon[|\psi_0\rangle\langle\psi_0|]=(1-q)|\psi_0\rangle\langle\psi_0|+q|\psi_1\rangle\langle\psi_1|,$$
where $p=|\langle\psi_1|e_1\rangle|^2$ and $q=|\langle\psi_0|e_0\rangle|^2$.
\begin{figure}
  \includegraphics[width=5cm]{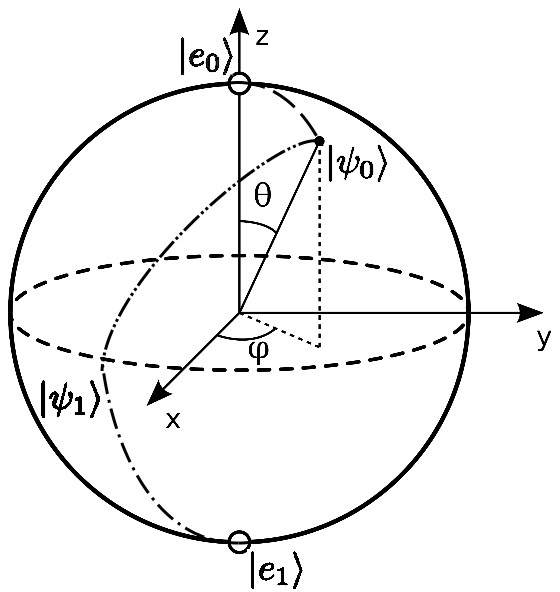}\\
  \caption{Bloch sphere}\label{BSR}
\end{figure}
Now recalling that the angle $A$ (see equation (\ref{eq.angle})) is a distance measure, we obtain that

 \begin{tabular}{rcl}
      $\frac{\pi}{2}$ & $=A(|e_0\rangle\langle e_0|,|e_1\rangle\langle e_1|)\leq $&  \\
    & $\leq A(|e_0\rangle\langle e_0|,|\psi_0\rangle\langle \psi_0|)+$& $A(|\psi_0\rangle\langle \psi_0|,|\psi_1\rangle\langle \psi_1|)+$ \\
    &  & $A(|\psi_1\rangle\langle \psi_1|,|e_1\rangle\langle e_1|)$,\\
 \end{tabular}

 that gives
$$\frac{\pi}{2}\leq\arccos\sqrt{p}+\arccos{|\langle \psi_0|\psi_1\rangle|}+\arccos\sqrt{q}.$$
Note that if $\langle \psi_0|\psi_1\rangle\neq0$ then $p$ and/or $q$ are  necessarily less then $1$.
Alternatively, the same result can be  proved making resort to the Bloch sphere representation. In fact $2 A(|a\rangle\langle a|,|b\rangle\langle b|)$ is the arc  between the two pure states $|a\rangle,|b\rangle$ in the Bloch sphere. In particular to minimize the probability of error we need to have the four states $|e_0\rangle,|e_1\rangle,|\psi_0\rangle$ and $|\psi_1\rangle$ on the same diameter. The zero error case is obtained only when  $|\psi_0\rangle$ and $|\psi_1\rangle$ are orthogonal.
This last observation proves  that  an inevitable error exists connected to the use of non orthogonal basis for the computation.
We stress finally that these results suggest to   convert the equation (\ref{eq.tau})  into an informational inequality that connects physical quantity:
\begin{equation}\label{eq.ine}
    \Delta t\geq\frac{2\hbar}{\Delta E}\varepsilon,
\end{equation}
where $\Delta t$ represents the time needed to perform the transformation, $\Delta E$ is the difference between the maximal and minimal energy values and $\varepsilon$ is a parameter that characterize the efficiency of the computational process (that depends on the geometry of the problem).

\section*{Conclusions}

In this paper we have analyzed the Quantum Brachistochrone problem in Hermitian and quasi Hermitian quantum mechanics, focusing our attention on the  physical realizability of these  processes.
In particular this study shows that the violation for the  bound for the minimal time obtained in  quasi Hermitian quantum mechanics is only apparent, and can be justified by geometrical arguments. A simulation of PH dynamics by Hermitian open dynamics was analyzed in sections \ref{sec.4.4} and \ref{sec.4.5}. This possibility is physically realizable but it has an energy and efficiency cost.
In section \ref{sec.4.6} these results applied to simple quantum gates, suggest that equation
(\ref{eq.tau}) has an informational meaning that is quasi Hermitian invariant. In fact (\ref{eq.ine}) gives an explicit connection between efficiency, energy and time needed to do a computational process.

\section*{Acknowledgments}

I am grateful to prof. L. Solombrino for helpful discussions and suggestions.
\bibliography{Some_Remarks_on_Quantum_Brachistochrone}

\end{document}